\documentclass[12pt, a4paper]{article}
\usepackage{graphicx}

\begin{document}
\title{First-order super-radiant phase transitions in a multi-qubit--cavity system} \author{Chiu Fan Lee\thanks{c.lee1@physics.ox.ac.uk}
\ and \ Neil F. Johnson\thanks{n.johnson@physics.ox.ac.uk}
\\
\\ Centre for Quantum Computation and Physics Department \\ Clarendon Laboratory, Oxford University \\ Parks Road, Oxford OX1 3PU, U.K.}

\maketitle

\abstract{We predict the existence of novel first-order phase transitions in a general class of multi-qubit-cavity systems. Apart from atomic systems, the associated super-radiant phase transition should be observable in a variety of solid-state experimental systems, including the technologically important case of interacting quantum dots coupled to an optical cavity mode.}

\newpage

Phase transitions in quantum systems are of great interest to the solid-state and atomic communities \cite{Kad,Sac99}, and have even caught the attention of the quantum information community in connection with entanglement \cite{qip,nielsen}. Most of the focus within the solid-state community has been on phase transitions in electronic systems such as low-dimensional magnets \cite{Kad,Sac99} while in atomic physics there has been much interest in phase transitions in cold atom gases and in atoms coupled to a cavity. In particular, a second-order phase transition, from normal to superradiance, is known to arise in the Dicke model which considers $N$ two-state atoms (i.e. `spins' or `qubits' \cite{qip,nielsen}) coupled to an electromagnetic field (i.e. bosonic cavity mode) \cite{Dic54,HL73,WH73}. The Dicke model itself has been studied within the atomic physics community for fifty years, but has recently caught the attention of solid-state physicists working on arrays of quantum dots, Josephson junctions, and magnetoplasmas \cite{CMP}. Its extension to quantum chaos \cite{chaos}, quantum information \cite{entanglement} and other exactly solvable models has also been considered recently \cite{solvable}.

Despite its obvious appeal, the Dicke model ignores interactions between the `spins'. In atomic systems where each `spin' is an atom, this is arguably an acceptable approximation if the atoms are neutral {\em and} the atom-atom separation $d\gg a$ where $a$ is the atomic diameter. However there are several reasons why this approximation is unlikely to be valid in typical solid-state systems. First, the `spin' can be represented by any nanostructure (e.g. quantum dot) possessing two well-defined energy levels, yet such nanostructures are not typically neutral. Hence there will in general be a short-ranged (due to screening) electrostatic interaction between neighbouring nanostructures.
Second, even if each nanostructure is neutral, the typical separation $d$ between nanostructures in fabricated arrays is similar to the size of the individual nanostructures themselves ($10^2-10^3$A).
Hence neutral systems such as excitonic
quantum dots will still have a significant interaction between nearest neighbors \cite{us}.

Motivated by the experimental relevance of `spin--spin' interactions, we introduce and analyze a generalized Dicke Hamiltonian which is relevant to current experimental setups in both the solid-state and atomic communities \cite{expt}. We show that the presence of transverse spin--spin coupling terms, leads to novel first-order phase transitions associated with super-radiance in the bosonic cavity field. A technologically important example within the solid-state community would be an array of quantum dots coupled to an optical mode. This mode could arise from an optical cavity, or a defect mode in a photonic band gap material \cite{expt}.
However we emphasize that the $N$ `spins' may correspond to  {\em any} two-level system, including superconducting qubits and atoms \cite{CMP,expt}. The bosonic field is then any field to which the corresponding spins couple \cite{CMP,expt}.
Apart from the experimental prediction of novel phase transitions, our work also provides an interesting generalization of the well-known Dicke model.

Just as the themodynamic results for the original Dicke model turned out to be valid for a wider class of Dicke-like Hamiltonians \cite{WH73}, the results we present are actually valid for a wider class of Hamiltonian incorporating spin--spin and spin--boson interactions \cite{unpub}.
However for simplicity, we will focus here on a straightforward example:
\begin{eqnarray} H&=&a^\dag a +
  \sum_{j=1}^{N} \left\{ \frac{\lambda}{2 \sqrt{N}}
  (a +a^\dag)(\sigma^+_j+
  \sigma^-_j) + \frac{\epsilon}{2} \sigma^Z_j  - J \sigma^Y_j \cdot \sigma^Y_{j+1} \right\} \\ &=&a^\dag a +
  \sum_{j=1}^{N} \left\{\frac{\lambda}{2 \sqrt{N}}
  (a +a^\dag)\sigma^X_j + \frac{\epsilon}{2} \sigma^Z_j
  - J \sigma^Y_j \cdot \sigma^Y_{j+1} \right\} \ .
\end{eqnarray}
Following the discussions above, the experimental spin--spin interactions are likely to be short-ranged hence only nearest-neighbor interactions are included in $H$. The operators in Eqs. 1 and 2 have their usual, standard meanings. To solve for the thermodynamic properties of $H$, we introduce Glauber coherent states $| \alpha \rangle $, which have the following properties \cite{WH73}:
$a|\alpha \rangle = \alpha | \alpha \rangle$,
  $\langle \alpha | a^\dag = \langle \alpha | \alpha^*$, and $\int \frac{d {\rm Re} (\alpha) d {\rm Im} (\alpha) }{\pi}  |\alpha \rangle \langle \alpha| =1$. In terms of this basis, the canonical partition function can be written as:
\begin{equation} Z(N,T)=\sum_{\bf s} \int \frac{d {\rm Re} (\alpha) d {\rm Im} (\alpha) }{\pi} \langle {\bf s} | \langle \alpha| e^{-\beta H} | \alpha \rangle | {\bf s} \rangle \end{equation} We adopt the following assumptions as in Ref. 
\cite{WH73}:
\begin{enumerate}
\item
$a/\sqrt{N}$ and $a^\dag/ \sqrt{N}$ exist as $N \rightarrow \infty$; \item $\lim_{N \rightarrow \infty} \lim_{R \rightarrow \infty} \sum_{r=0}^R \frac{(-\beta H_N)^r}{r!}$ can be interchanged \end{enumerate} We then find \begin{equation}
Z(N,T) = \int \frac{d^2 \alpha}{\pi} e^{-\beta|\alpha|^2}({\rm Tr} e^{-\beta H'})^{N} \end{equation} where \begin{eqnarray} H' &=& \sum_{j=1}^N \left\{
  \frac{\lambda {\rm Re}( \alpha)}{\sqrt{N}} \sigma^X_j
  + \frac{\epsilon}{2} \sigma^Z_j  -J \sigma^Y_j \cdot \sigma^Y_{j+1} \right\}\\ &=& -J\sum_{j=1}^N
  \left\{ \sqrt{\left( \frac{\lambda {\rm Re}(
\alpha)}{J\sqrt{N}}\right)^2
+\left( \frac{\epsilon}{2J}\right)^2} \sigma^Z_j
  +   \sigma^Y_j \cdot \sigma^Y_{j+1} \right\} \\ &=&
\label{epsilon1}
\sum_{k=1}^N \xi_k(\alpha) (\gamma_k^\dag \gamma_k -\frac{1}{2} ) \end{eqnarray} with \begin{eqnarray} \label{epsilon2} \xi_k (\alpha) &=& 2J \sqrt{1+(g(\alpha))^2+ 2g(\alpha) } \\ g(\alpha) &=& \sqrt{\left( \frac{\lambda {\rm Re} ( \alpha)}{J\sqrt{N}}\right)^2
	+\left( \frac{\epsilon}{2J}\right)^2 } \ .
\end{eqnarray} Equation 6 follows from a rotation about the $y$-axis, while Eqs. \ref{epsilon1} and \ref{epsilon2} are derived in Ref. \cite{Sac99} for example. With $H'$ diagonalized, $Z(N,T)$ becomes \begin{eqnarray} && \int \frac{d^2 \alpha}{\pi} e^{-\beta |\alpha|^2} \left\{ \sum_{k=1}^N \sum_{n_k =0}^1 \langle 0 |\gamma_1^{n_1} \cdots \gamma_N^{n_k} e^{-\beta H'} (\gamma^\dag_1)^{n_1} \cdots (\gamma^\dag_N)^{n_k} |0 \rangle \right\} \nonumber \\ &=& \int \frac{d^2 \alpha}{\pi} e^{-\beta \left(|\alpha|^2- \sum_k \xi_k ({\rm Re} (\alpha)) / 2 \right)} \left\{ \prod_{k=1}^N \left(1+ e^{-\beta \xi_k  ({\rm Re} (\alpha)) }
\right) \right\} \\  &=&
\frac{1}{\sqrt{ \beta \pi}} \int  dw e^{-\beta \left(w^2- \sum_k \xi_k (w) / 2 \right)} \left\{ \prod_{k=1}^N \left(1+ e^{-\beta \xi_k  (w) } \right) \right\} \end{eqnarray} where $w= {\rm Re}(\alpha)$.
Letting $x=w/\sqrt{N}$ and writing $\sum_{k=1}^N$ as $\frac{N}{2 \pi} \int_0^{2\pi} dk$, $Z(N,T)$ becomes \begin{equation} \sqrt{\frac{N}{\beta \pi}} \int_{-\infty}^\infty dx \left\{ e^{- \beta x^2 + I(x)} \right\}^N \end{equation} where \begin{eqnarray} I(x)&=& \frac{1}{2 \pi}
\int_0^{2 \pi} dk \left\{
\log \left[ \cosh \left(
\frac{\beta}{2} \xi_k(x)  \right) \right] + \log (2) \right\} \end{eqnarray} and \begin{equation} \xi_k (x) = 2J \sqrt{1+(g(x))^2+ 2 g(x) \cos k } \ .
\end{equation}
From here on, we will omit the $\log (2)$ term in $I(x)$ since it only contributes an overall factor to $Z(N,T)$.

Laplace's method now tells
us that
\begin{equation}
\label{laplace1} Z(N,T) \propto \max_{-\infty \leq x \leq \infty} \exp \left\{ N [-\beta x^2 + I(x)] \right\}.
\end{equation}
Denoting $[-\beta x^2 + I(x)] $ by $\Omega(x)$, we recall that the super-radiant phase corresponds to $\Omega(x)$ having its maximum at a non-zero $x$ \cite{WH73}. If there is no transverse field, i.e., if $J=0$, and the temperature is fixed, then the maximum of $\Omega(x)$ will split continuously into two maxima symmetric about the origin as $\lambda^2$ increases.
Hence the process is a
continuous phase transition.

However the case of non-zero $J$ is qualitatively different from $J=0$. 
As a result
of the frustration induced by the tranverse nearest-neighbor couplings, {\it there are regions where the super-radiant phase transition becomes first-order.} This phenomenon of first-order phase transitions is revealed by considering the functional shape of $I(x)$.
Numerical simulations show
that $I(x)$ can have one, two or three local maxima, as can be deduced from the behavior of $\Omega(x)$ in Fig. 1.

Figure 2 shows the regions
in which these three cases appear for fixed $J$ and $\beta$.
These regions are determined by considering the second derivatives of $I(x)$ for various $\lambda, J, \epsilon$ and $\beta$. Figure 3 plots the maximizer of $\Omega(x)$ with $\lambda$ fixed at a value of 1.3. Referring to Fig. 
2, we
see that in the region with three maxima, e.g. when $\epsilon \leq 1$, the non-zero local maximizers are dominant and hence the super-radiant state appears.
As
$\epsilon$ increases, these two local maximizers converge to zero  and the system is no longer super-radiant. This is no longer the case if $J$ is increased slightly, e.g. to 0.56. In this case, $\Omega$ has a global maximum when $\epsilon$ is small; however as $\epsilon$ increases, the non-zero local maxima become dominant and as a result, a first-order phase transition occurs.
We note that the barriers between the wells are infinite in the thermodynamic limit, hence we expect that the sub-radiant state is metastable as $\epsilon$ increases. This observation also suggests the phenomenon of hysteresis, which awaits experimental validation.

The relevant parameter ranges $J\sim 0.1-0.6$, $\beta=100$, $\epsilon\sim 1$ and $\lambda\sim 0.5-1.5$ correspond to realistic scaled parameter values for both nanostructure-cavity and atom-cavity systems \cite{us,expt,Koch}. For example in semiconductor quantum dot systems, $\epsilon$ is of order of the underlying bulk bandgap and hence can easily be chosen to range from $1.5$eV in GaAs-like systems down to $0.1$eV in narrow-gap semiconductors. Meanwhile $J$ is determined independently by the interdot separation which can be chosen to range from $10^2-10^3$ Angstroms, hence $J$ can be engineered to be of the order of $\sim 10-10^2$ meV. The value of $\lambda$ can also be chosen independently, according to the strength of the coupling to the cavity mode. Both weak and strong coupling regimes are accessible using current nanotechnology.

In conclusion, we have shown that the experimentally relevant spin-spin interaction transforms the nature of the super-radiant phase transition in the Dicke model. Our results highlight the importance of spin-spin coupling terms in spin-boson systems and open up many questions regarding the possible competition between the sub-radiant and super-radiant states in experimental systems.

C.F.L. thanks University College for a Research Fellowship and NSERC
(Canada) for partial financial support. N.F.J. thanks the LINK-DTI project, and is very grateful to Luis Quiroga and Alexandra Olaya-Castro for continuing discussions.

\begin{figure}
\caption{
The function $\Omega(x)$, which governs the phase transitions in the 
system, shown
for various values of $J$ and $\epsilon$ with $\lambda$ fixed at 1.3.
The $y$-axes are shifted for each curve so that all
curves are zero at the origin.}
\begin{center}
\includegraphics{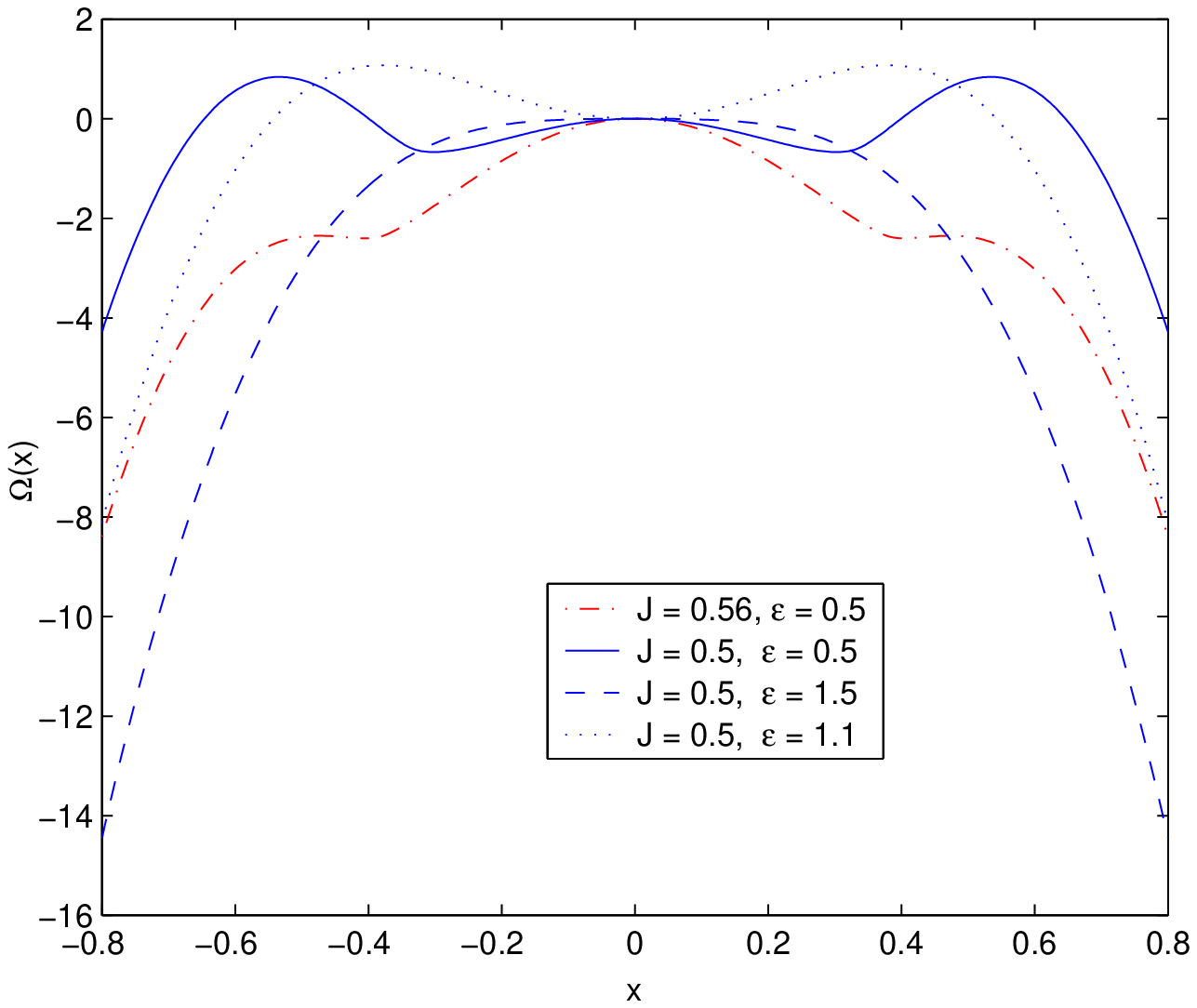}
\end{center}
\end{figure}

\begin{figure}
\label{J_fixed}
\caption{
Regimes of behavior of the function $\Omega(x)$ which governs the phase
transitions in the system.
The red broken lines represent various values of
$J$ ($J=0.1,0.2,0.3,0.4,0.5,0.6$ from bottom to top). Above the red 
broken line and
to the left of the blue solid line signifies the region where
$\Omega(x)$ has two local maxima. Below the red broken line and to the 
left of
the blue solid line, signifies the region where
$\Omega(x)$ has three local maxima. $\Omega(x)$ has only one maximum to 
the
right of the blue solid line. $\beta$ is taken to be 100. }
\begin{center}
\includegraphics{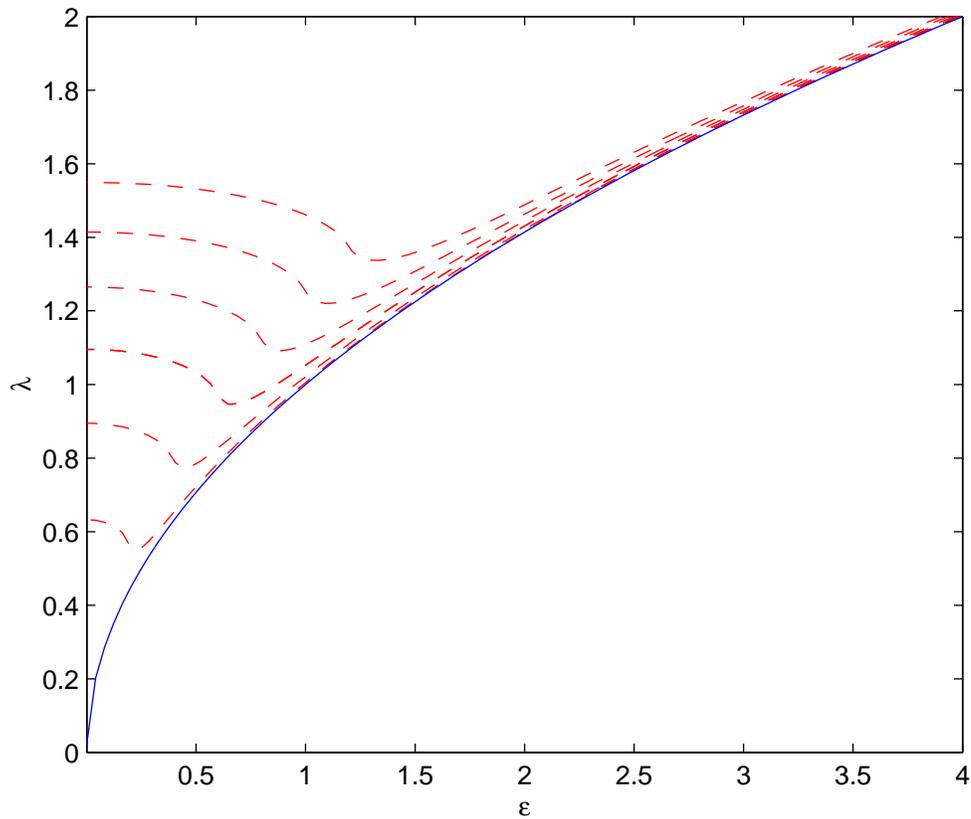}
\end{center}
\end{figure}

\begin{figure}
\caption{
Plot of the maximizer of the function $\Omega(x)$, which governs the 
phase transitions
in the system, as a function of $J$ and $\epsilon$.
$\lambda$ and $\beta$ are fixed to 1.3 and 100 respectively.}
\begin{center}
\includegraphics{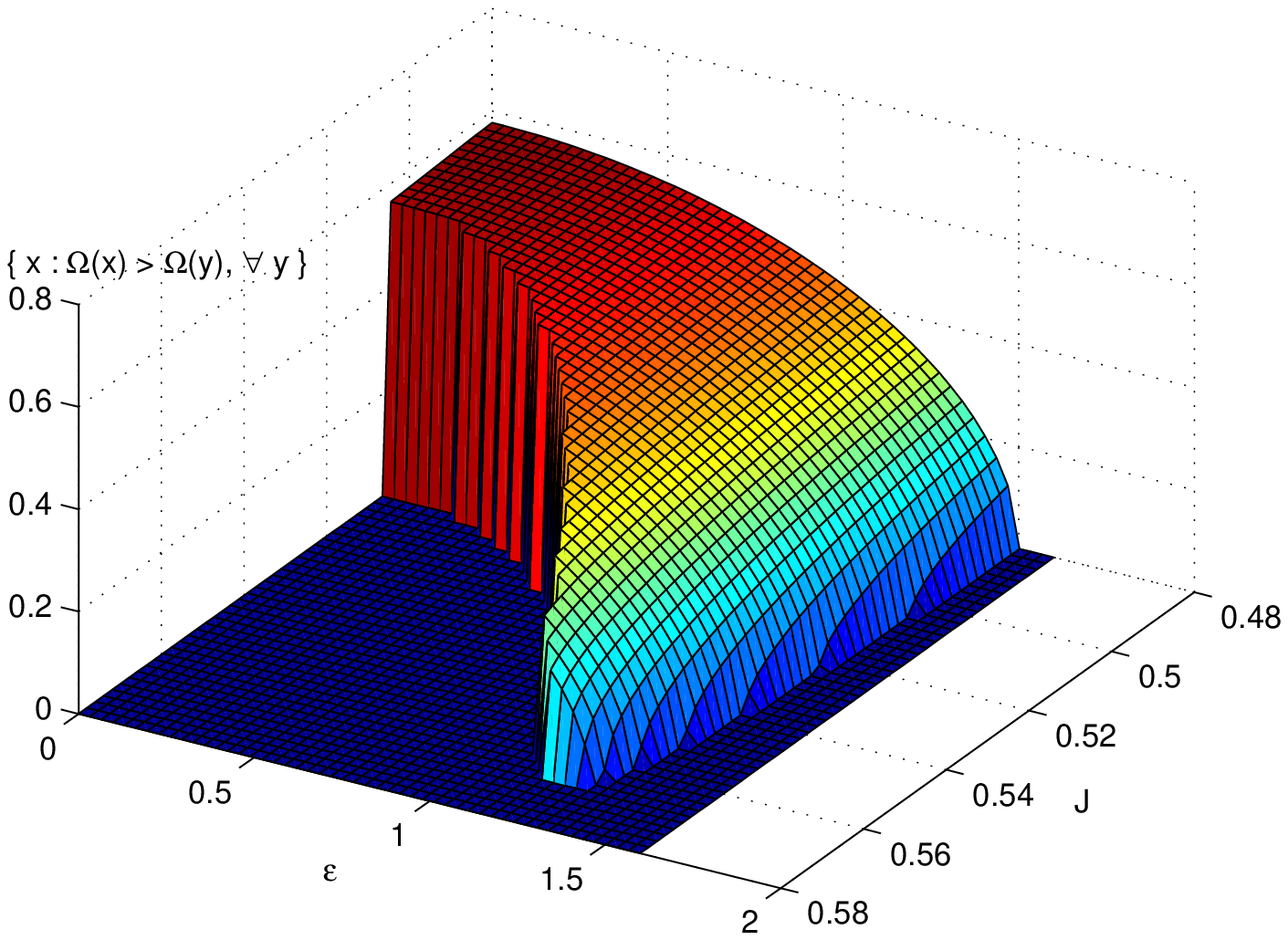}
\end{center}
\end{figure}


\newpage
\begin{thebibliography}{99}

\bibitem{Kad} L.P. Kadanoff, \emph{Statistical Physics} (World 
Scientific,
Singapore, 2000).

\bibitem{Sac99} S. Sachdev, \emph{Quantum Phase Transitions} (Cambridge
University Press, Cambridge, 1999).

\bibitem{qip} T.J. Osborne and M.A. Nielsen, Phys. Rev. A {\bf 66}, 
032110 (2002);
R. Somma, G. Ortiz, H. Barnum, E. Knill, and L. Viola, quant-ph/0403035.

\bibitem{nielsen} M.A. Nielsen and I.L. Chuang,
{\it Quantum Computation and Quantum
Information} (Cambridge  University Press, Cambridge, 2002).

\bibitem{Dic54} R.H. Dicke, Phys. Rev. {\bf 170}, 379 (1954).

\bibitem{HL73} K. Hepp and E.H. Lieb, Ann. Phys. (N.Y.) {\bf 76}, 360 
(1973).

\bibitem{WH73} Y.K. Wang and F.T. Hioe, Phys. Rev. A {\bf 7}, 831 
(1973); F.T.
Hioe, Phys. Rev. A {\bf 8}, 1440 (1973).

\bibitem{CMP}
T. Vorrath and  T. Brandes, Phys. Rev. B {\bf 68}, 035309 (2003);
W.A. Al-Saidi and  D. Stroud, Phys. Rev. B {\bf 65}, 224512 (2002); X. 
Zou,  K.
Pahlke and  W. Mathis, quant-ph/0201011; S. Raghavan,  H. Pu,  P. 
Meystre and
N.P. Bigelow, cond-mat/0010140; N. Nayak,  A.S. Majumdar and  V. 
Bartzis, J.
Nonlinear Optics {\bf 24}, 319 (2000); T. Brandes,  J. Inoue and  A. 
Shimizu,
cond-mat/9908448 and cond-mat/9908447.

\bibitem{chaos} C. Emary and T. Brandes, Phys. Rev. Lett. {\bf 90}, 
044101
(2003); Phys. Rev. E {\bf 67}, 066203 (2003).

\bibitem{entanglement} N. Lambert, C. Emary and T. Brandes, Phys. Rev. 
Lett.
{\bf 92}, 073602 (2004);  S. Schneider and G.J. Milburn, 
quant-ph/0112042;
G. Ramon,  C. Brif and  A. Mann,
Phys. Rev. A {\bf 58}, 2506 (1998);  A. Messikh,  Z. Ficek and M.R.B.
Wahiddin, quant-ph/0303100; A. Olaya-Castro, N.F. Johnson, L. Quiroga, 
J. Phys. B
(in press).

\bibitem{solvable}
C. Emary and T. Brandes, quant-ph/0401029;
S. Mancini,  P. Tombesi and V.I. Man'ko, quant-ph/9806034.

\bibitem{us} L. Quiroga and N.F.
Johnson, Phys. Rev. Lett. {\bf 83}, 2270 (1999); J.H. Reina, L. Quiroga 
and N.F.
Johnson, Phys. Rev. A {\bf 62}, 012305 (2000).

\bibitem{expt} E. Hagley et al., Phys. Rev. Lett. {\bf 79}, 1 (1997); A.
Rauschenbeutel et al., Science {\bf 288}, 2024 (2000);
A. Imamoglu et al., Phys. Rev. Lett. {\bf 83}, 4204 (1999);  S.M. 
Dutra, P.L. Knight and H.
Moya-Cessa, Phys. Rev. A {\bf 49}, 1993 (1994); Y. Yamamoto and R. 
Slusher,
Physics Today, June (1993), p. 66; D.K. Young, L. Zhang, D.D. Awschalom 
and E.L.
Hu, Phys. Rev. B {\bf 66}, 081307 (2002); G.S. Solomon, M. Pelton and Y.
Yamamoto, Phys. Rev. Lett. {\bf 86}, 3903 (2001); B. Moller, M.V. 
Artemyev, U. Woggon and R. Wannemacher, 
Appl. Phys. Lett. {\bf 80}, 3253 (2002); A. Olaya-Castro, N.F. 
Johnson
and L. Quiroga, quant-ph/0311111; N. F. Johnson, J. Phys. Condens. 
Matter 7, 965
(1995). For molecular biological systems, see A.J. Berglund, A.C.
Doherty and H. Mabuchi, Phys. Rev. Lett. {\bf 89}, 068101 (2002). For a
discussion of photonic band gap materials, see P.M. Hui and
N.F. Johnson, Solid State Physics, Vol. 49, ed. by H. Ehrenreich and F. 
Spaepen
(Academic Press, New York, 1995).

\bibitem{unpub} C.F. Lee and N.F. Johnson, unpublished.

\bibitem{Koch} H. Haug and S.W. Koch, {\em Quantum theory of the 
optical and
electronic properties of semiconductors} (World Scientific, Singapore, 
2004).



\end{thebibliography}
\end{document}